\documentstyle[epsf]{elsart}
\newcommand{\LSGO}{LaSrGaO$_4$}
\newcommand{\tc}{$T_c$}
\newcommand{\cm}{cm$^{-1}$}
\newcommand{\YBCO}{YBa$_2$Cu$_3$O$_{7-\delta}$}

\begin{document}
\begin{frontmatter}
\title{Temperature Dependence of the FIR Reflectance of \LSGO}
\author{A.W.~McConnell\thanksref{new}}
\and
\author{ T.~Timusk} 

\address{Department of Physics and Astronomy, McMaster University, 
 Hamilton, Ontario, Canada L8S~4M1}

\author{A.~Dabkowski and H.A.~Dabkowska}
\address{Brockhouse Institute of Materials Research, McMaster University, 
 Hamilton, Ontario, Canada L8S~4M1}

\date{10 Sept. 1997}

\begin{abstract}

The reflectance of single crystal \LSGO\  has been measured from $\approx$~50
to 40000~\cm\ along the {\em a} and{\em c} axis. The optical properties have
been calculated from a Kramers-Kronig analysis of the reflectance for both
polarization. The reflectance curves have been fit using a product
of Lorentzian oscillators. 


\end{abstract}


\thanks[new]{ Present address: Department Of Physics, Simon Fraser University, Burnaby,
B.C.,Canada V5A~1S6. }

\end{frontmatter}
\epsfclipon
%
%
%
\section*{Introduction}
The importance of film growth in the field of high temperature HTSC
superconductors has lead to a search for new substrates with 
properties optimized for various applications. The substrates have to 
be compatible with the film both in physical properties such as 
lattice parameter, thermal expansion coefficient as well as chemical 
properties, for example the  avoidance of certain elements such as 
silicon which can diffuse from the substrate to the film. The use of 
the film in passive microwave devices, or to fabricate Josephson 
junctions further restrict the choice of substrate materials.
Popular substrate  materials for HTSC films include LaAlO$_3$, SrTiO$_3$, 
NdGaO$_3$, MgO, with the majority of the work being done on 
\YBCO. Recently several groups have begun using the crystal 
substrate, \LSGO, which has some promising properties for the 
production of microwave devices and Josephson junctions.

\LSGO\  has a tetragonal K$_2$NiF$_4$ structure with the La and Sr distributed
statistically on the K sites.\cite{ruter90} The space group of this
structure is I4/{\em mmm} (D$^{17}_{4h}$). Seven infrared active modes are
predicted by group theory\cite{burns88}; four E$_u$ modes perpendicular to
the {\em c} axis and three A$_{2u}$ modes parallel to the {\em
c} axis.\cite{tajima91} The lattice parameters of \LSGO\ are {\em a}=
3.84~\AA\ and {\em c}=12.86~\AA\ at room temperature with the {\em a} axis
being a close match to the {\em a} and {\em b}-axes of \YBCO. The room
temperature dielectric constant at 10~GHz is $\epsilon_1=22$ which compares
favorably with that of LaAlO$_3$. The thermal expansion coefficient along
the {\em a} axis is 10 ppm/$^\circ$C providing a reasonable match to
that of \YBCO\ (12.6 ppm/$^\circ$C).

Hontsu {\em et al.} and others have shown that \LSGO\ can be used to 
grow good quality \YBCO\ thin films.\cite{nakamura93,hontsu91} A 
recent comparison study indicates that the \YBCO\ films grown on 
\LSGO\ were of comparable quality in transition temperature, critical 
current, and surface morphology, to the films grown on 
LaAlO$_3$.\cite{mcconnell94} 
Most importantly for microwave applications, \LSGO does not suffer 
from any phase transition which can lead to twinning as does 
LaAlO$_3$. It has been reported that \LSGO\ has 
been used to produce \YBCO\ {\em a} axis films which display a high 
degree of in-plane alignment of the {\em c} 
axis.\cite{ito94,hontsu92,suzuki93,mukaida93} This is of interest 
because  good quality {\em a} axis films may be used in the 
manufacture of Josephson junctions; they 
are also convenient for studies of the properties of \YBCO\ thin 
films. \LSGO\ has also been used for the growth of {\em a} axis 
aligned Bi$_2$Sr$_2$CuO$_x$ films.\cite{ishibashi92} A number of 
studies of junctions utilizing \LSGO\ as a substrate or as a barrier 
layer have been published, pointing to its potential in device 
manufacturing.\cite{wen91,lew94,matsui92,hontsu94} One study of a 
variety of substrates suggests that \LSGO\ may allow the manufacture 
of a microwave resonator and filter with a temperature independent 
center frequency.\cite{konaka91} 

A study of the infrared properties of the substrate material serves 
several goals.  One can form an estimate of the fundamental microwave 
properties of a dielectric material from an extrapolation of the low 
frequency portion of its infrared spectrum.\cite{petzelt93} Also, for 
fundamental studies of the properties of thin films it is often found 
that if the films are thinner than the microwave penetration depth the 
fields penetrate into the substrate and the measurements have to be 
corrected for substrate effects by including corrections that assume 
a knowledge of the optical constants of the substrate.  

Reflectance measurements polarized along the axes of thin films are of use
in comparing the results with those of single crystals and offer a ready
technique for characterization when single crystals of sufficient size are
difficult to produce. However, a knowledge of the optical properties of the
underlying substrate is critical to the interpretation of the spectrum
since it is a convolution of the optical properties of both the film and
the substrate. With this as a motivation, the optical properties of
\LSGO\ were measured along both {\em a} and {\em c} polarizations from
50~\cm\ to 40000~\cm.

%
%
\section*{Experimental}

The \LSGO\ crystals used were grown at the Brockhouse Institute of 
Materials Reserach at McMaster University  
using a Czochralski
growth technique described elsewhere.\cite{dabkowski93} The samples used
for optical measurements were cut in two orientations, (001) for {\em
a} axis measurements and (010) for {\em c} axis. The samples were polished
to an optically smooth finish and mounted on flat holders.

The far infrared (FIR) measurements were made on a Michelson interferometer
using an {\em in situ} gold coating technique to
determine the absolute reflectance.\cite{homes93} We estimate the 
uncertainty of $\pm$1\% in the overall reflectance level. Bolometer 
detectors at 1.4K and a 4.2K were used in the 50~\cm\ to 800~\cm 
frequency range, and an MCT 
detector was used from 700~\cm\ to 7000~\cm. The 
resolution was 4~\cm\ up to 800~\cm\  and 10~\cm\ between 1000~\cm\ 
to 7000~\cm. The samples were cooled by a controlled liquid helium 
flow through a cold finger. 

The remaining frequencies from 7000~\cm\ to 40000~\cm\ were measured using
a diffraction grating spectrometer using {\em in situ} deposition of gold
or aluminum. The resolution for this region was 200~\cm. The measurements
above 800~\cm\ were made at room temperature, since the samples displayed no
significant temperature dependence in the reflectance above this frequency.

\section*{Results}

\begin{figure}
\epsfxsize=5.5in \epsffile{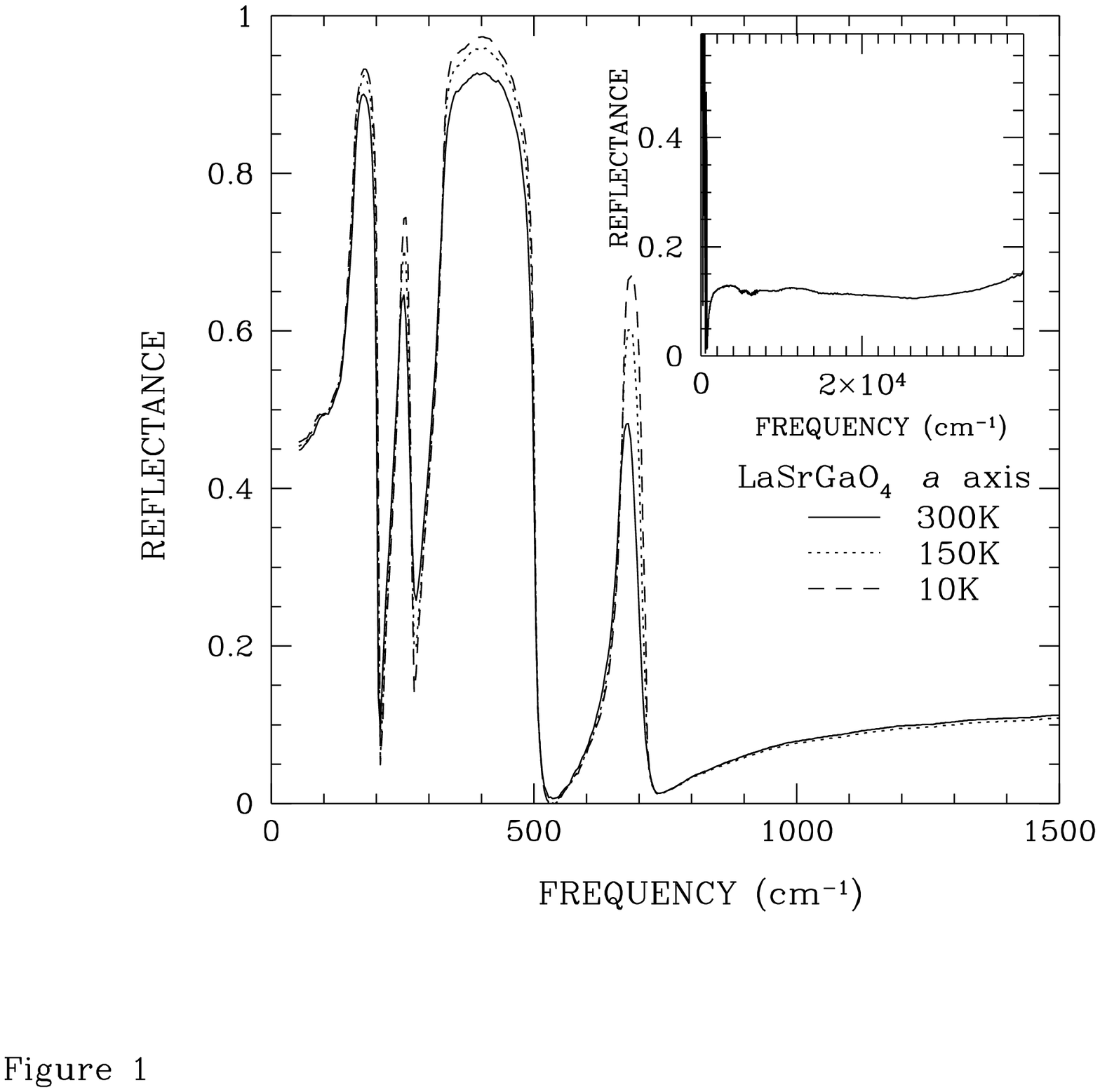}
\caption{\label{a-refl} Temperature dependence of the {\em a} axis
reflectance of \LSGO\  from 50~\cm to 1000~\cm at 300K (solid
lines), 150K (dotted line) and 10K (dashed line). The inset shows reflectance
spectrum for 300K out to 40000~\cm.}
\end{figure}

\begin{figure}
\epsfxsize=5.5in \epsffile{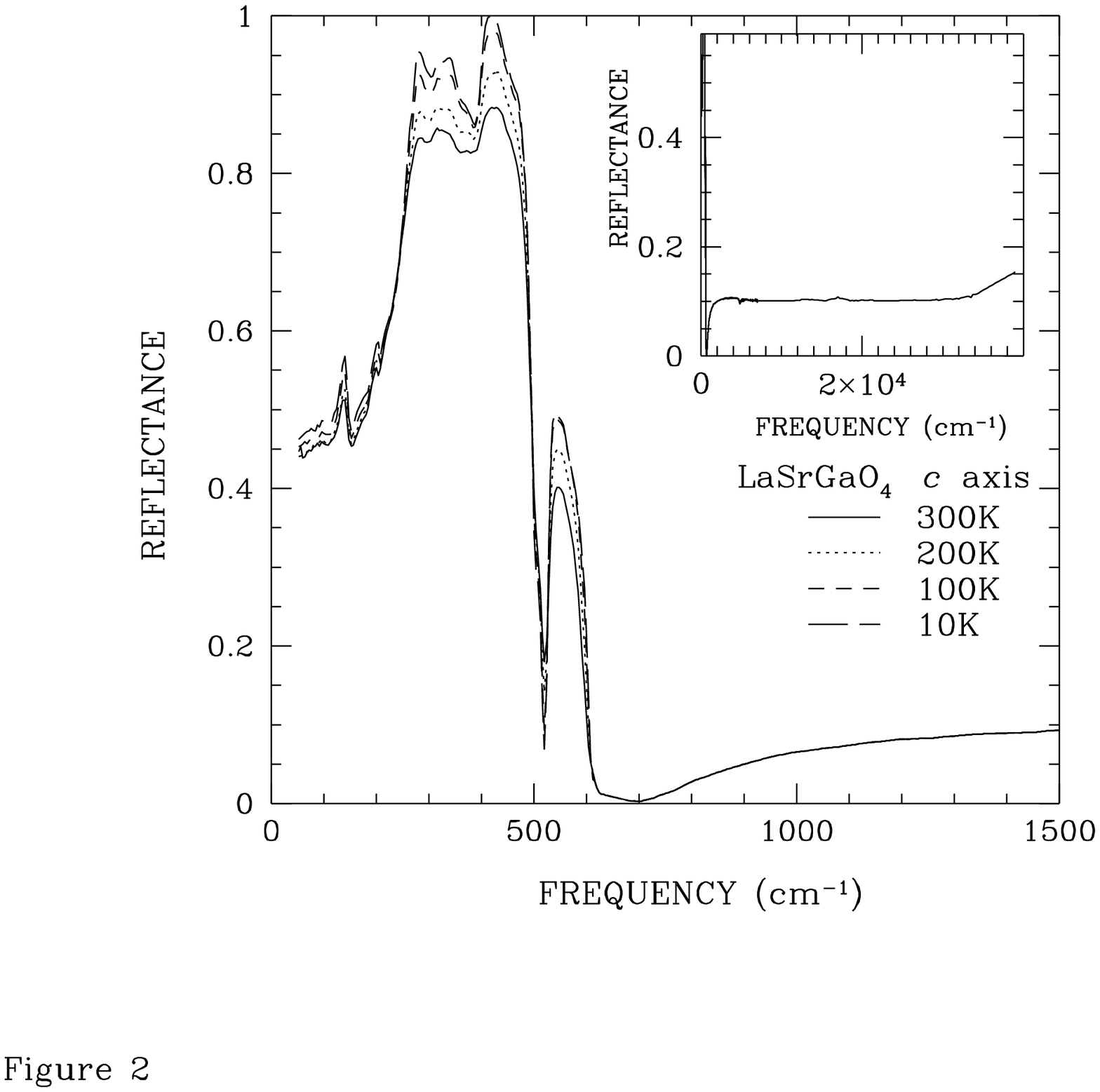}
\caption{\label{c-refl} Temperature dependence of the {\em c} axis
reflectance of \LSGO\  from 50~\cm to 1000~\cm at 300K (solid lines), 200K
(dotted line), 100K (short dashed line), and 10K (long dashed line).  The inset
shows reflectance spectrum for 300K out to 40000~\cm.}
\end{figure}

Figure~\ref{a-refl} shows the reflectance polarized along the {\em a} axis from 50 to
1500~\cm\ at 300~K, 150~K, and 10~K.  The inset shows the 300~K {\em a} axis reflectance
for the full range measured, 50 to 40000~\cm.  Figure~\ref{c-refl} shows the reflectance
polarized along the {\em c} axis from 50 to 1500~\cm \ for temperatures 300~K, 200~K,
100~K and 10~K.  The inset shows the 300~K {\em c} axis reflectance from 50 to 40000~\cm. 
The heights of the reflectance peaks can be seen to be increasing with decreasing
temperature in a monotonic fashion.

%

%
%

%
%

The optical properties of the sample can be extracted from the spectra
using Kramers-Kronig  transformations to calculate the phase,
$\phi$, of the reflectivity, $\tilde{r}=\sqrt{R}e^{\imath\phi}$. The 
optical constants $\epsilon_1$ and $\sigma_1$ are then calculated for 
each frequency using the following equations :
\begin{equation}
n={{1-R}\over{1+R-2\sqrt{R}\cos(\phi)}}
\end{equation}
\begin{equation}
k={{-2\sqrt{R}\sin(\phi)}\over{1+R-2\sqrt{R}\cos(\phi)}}
\end{equation} 
\begin{equation}
\tilde\epsilon={\epsilon_1+i{{4\pi\sigma_1}\over{\omega}}}
\end{equation}
where $\epsilon_1={(n^2-k^2)}$ and $\sigma_1={{2nk\omega}/{4\pi}}$.

\begin{figure}
\epsfxsize=5.5in \epsffile{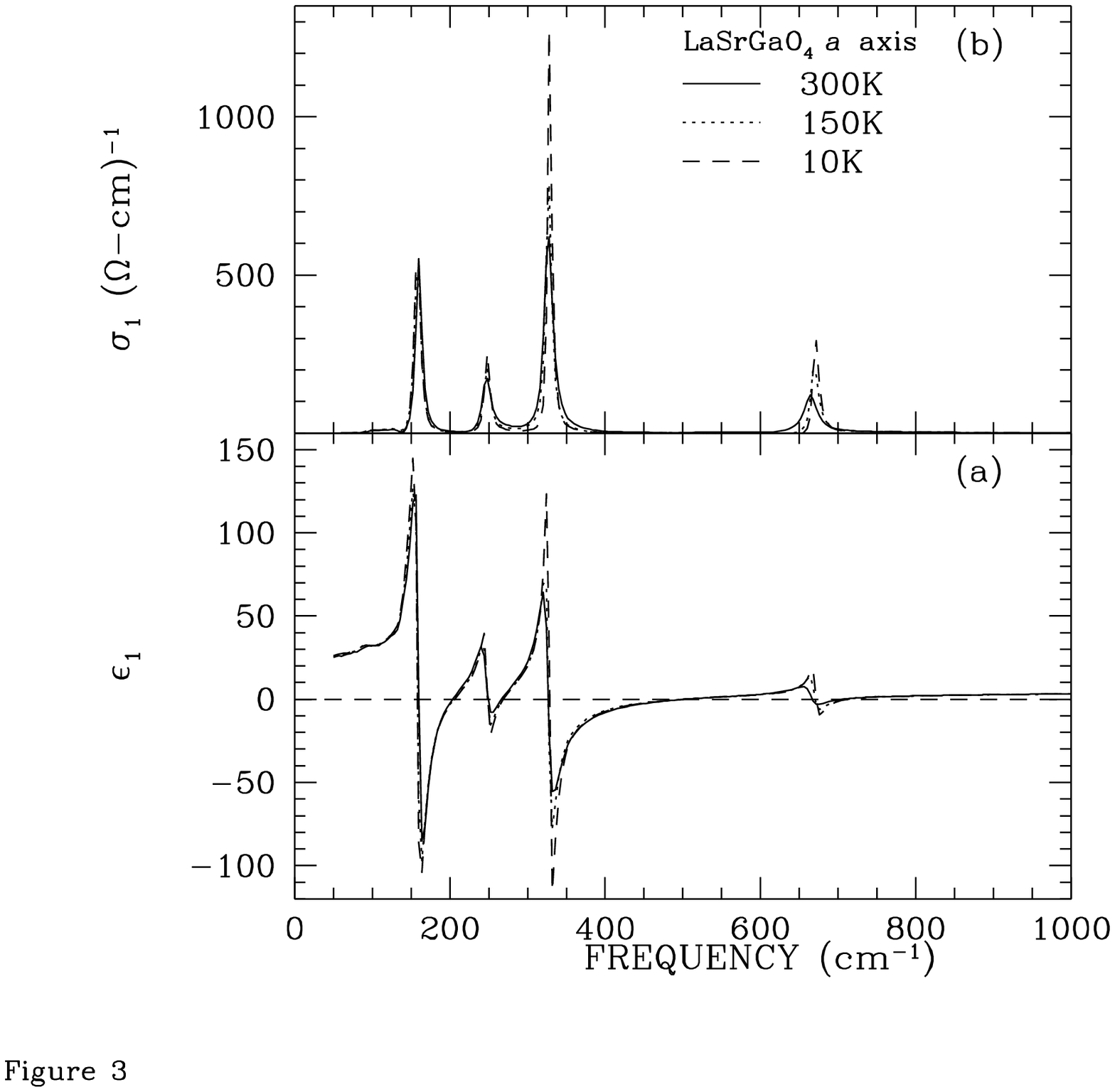}

\caption{\label{a-epsig}Temperature dependence of {\em a} axis dielectric
(a) and conductivity (b) of \LSGO. The temperatures shown are 300K (solid
line), 150K (dotted line), and 10K (dashed line).}
\end{figure}

\begin{figure}
\epsfxsize=5.5in \epsffile{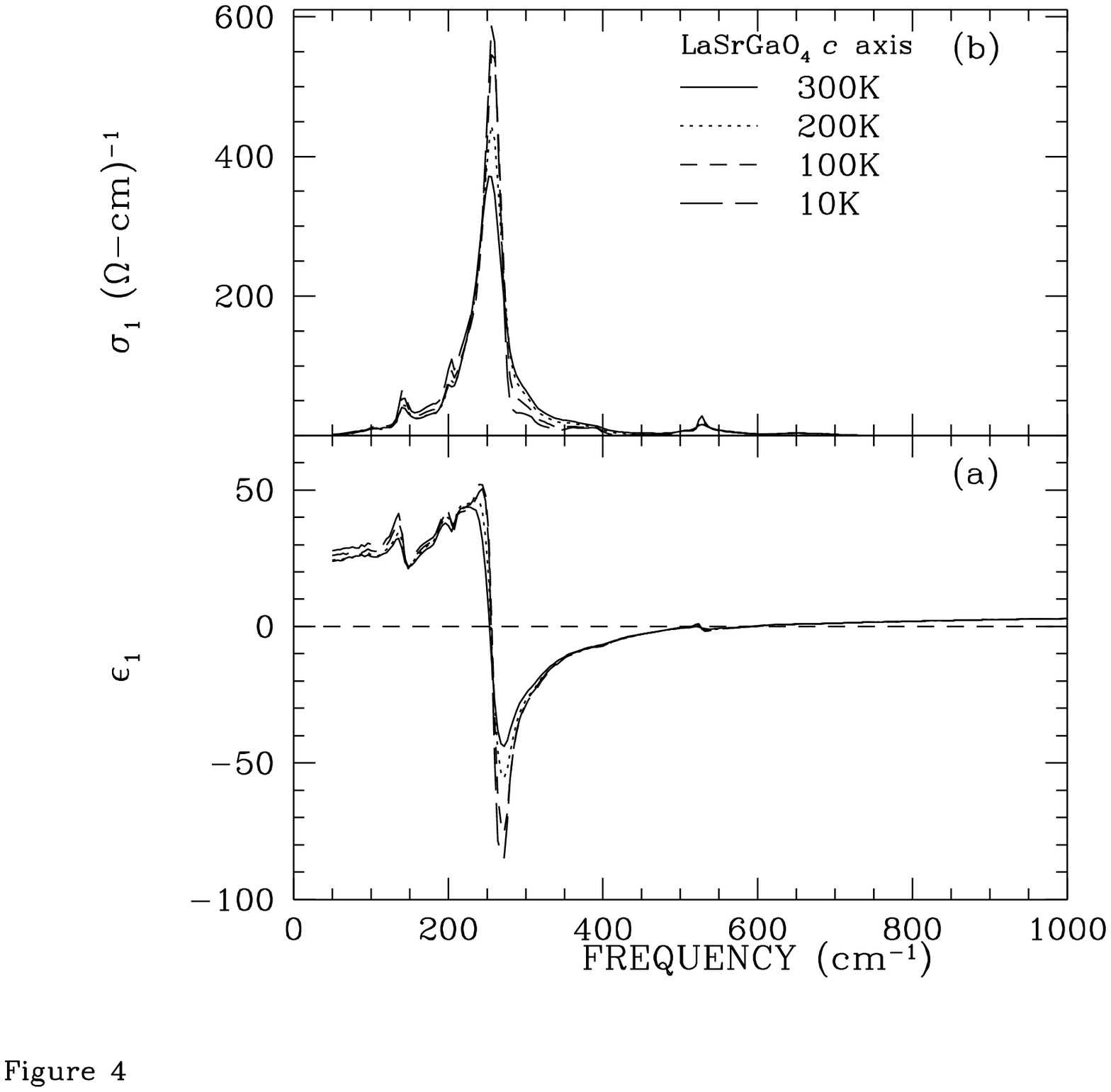}

\caption{\label{c-epsig}Temperature dependence of {\em c} axis dielectric
(a) and conductivity (b) of \LSGO. The temperatures shown are 300K (solid
line), 200K (dotted line), 100K (short dashed line), and 10K (long dashed
line). }
\end{figure}

Figure~\ref{a-epsig} and Figure~\ref{c-epsig} show the temperature
dependence of the real part of the dielectric function and the 
conductivity for the {\em a} axis and
{\em c} axis polarizations respectively. Due to the significant spectral
weight at frequencies above 40000~\cm, the calculation of these properties
is dependent on the choice of high-frequency extrapolation. As a result the
curves have only been presented up to 1000~\cm\ as below this frequency
there was only a weak dependence on the high-frequency extrapolation.  An
$\omega^{1/2}$ extrapolation out to $5\times10^5$~\cm\ followed by an
$\omega^{-4}$ form above this frequency were used to perform the
Kramers-Kronig calculation.\cite{wooten} A quadratic extrapolation to $R_0$
as $\omega\rightarrow 0$ was used for the low frequencies.  Looking at
Figure~\ref{a-epsig} (b), the phonon peaks in the conductivity all
increase with decreasing temperature with little shift in position, except
for the peak at about 675~\cm\ which can be seen to harden with decreasing
temperature.  None of the {\em c} axis phonon peaks in Figure~\ref{c-epsig}
(b) shift significantly.

%
%
%
%

In addition to performing the Kramers-Kronig calculation, the reflectance
data was fitted to a factorized form of the dielectric function
\cite{barker70} using a non-linear least-squares
technique. The form of the dielectric is as follows:

\begin{equation}
  \tilde\epsilon(\omega)=\epsilon_\infty \prod_j {{\omega^2_{LO,j}
  -\omega^2-i\omega\gamma_{LO,j}}\over{\omega^2_{TO,j}-\omega^2
  -i\omega\gamma_{TO,j}}},
\end{equation}
where $\omega_{LO,j}$, $\omega_{TO,j}$, $\gamma_{LO,j}$, and 
$\gamma_{TO,j}$ are the frequencies and widths of the {\em j}th LO
and TO modes, respectively.

The results of the fits are summarized in Tables~\ref{table1} and
\ref{table2} for the {\em a} and {\em c} axis data, respectively. The
fitting parameters have been limited by setting $\gamma_{LO,j}$, and
$\gamma_{TO,j}$ equal.  The oscillator strength for each mode,
${\omega_{pi}}^2$ can be calculated using\cite{genzel74}:
\begin{equation}
  \omega_{pi}^2=\epsilon_\infty(\omega^2_{LO,i}-\omega^2_{TO,i})
  \prod_{i\neq j} {{(\omega^2_{LO,j}-\omega^2_{TO,i})} \over
  {(\omega^2_{TO,j}-\omega^2_{TO,i})} },
  \ \ \ \ 
  (\gamma_{LO,i}=\gamma_{TO,i}).
\end{equation}
and the $\omega_{pi}$'s calculated are listed in Tables~\ref{table1} and
\ref{table2}. 

%

\begin{table}
\caption{\label{table1} Fitting parameters for the {\em a} axis reflectance
data utilizing the factorized product form. All parameters are in \cm \
except for $\epsilon_\infty$}
\begin{center}
\begin{tabular}{cccc}
 Oscillator  & 300K & 150K & 10K  \\
\hline
 $\omega_{TO,1}$ & 104.97& 103.64 & 101.31 \\
 $\omega_{LO,1}$ &106.37 & 105.77 & 103.62 \\
 $\gamma_{LO,1}$ &20.45  & 26.57  & 29.64  \\
 $\omega_{p1}$   &98.22  &120.55  & 124.73 \\
\hline
 $\omega_{TO,2}$ &160.84 & 159.50  & 159.04 \\
 $\omega_{LO,2}$ &203.34 & 205.14  & 205.94 \\
 $\gamma_{LO,2}$ &6.51   & 5.45    & 4.83   \\
 $\omega_{p2}$   &540.49 &543.65   & 551.26 \\
\hline
 $\omega_{TO,3}$ &246.73 & 247.17  & 247.80 \\
 $\omega_{LO,3}$ &269.09 & 269.33  & 269.10 \\
 $\gamma_{LO,3}$ &15.20  & 10.89   & 8.01   \\
 $\omega_{p3}$   &372.99 & 355.28  & 348.26 \\
\hline
 $\omega_{TO,4}$ &327.43 & 327.99  & 328.91 \\
 $\omega_{LO,4}$ &502.38 & 501.53  & 501.14 \\
 $\gamma_{LO,4}$ &10.41  & 6.96    & 5.55   \\
 $\omega_{p4}$   &687.11 & 668.27  & 676.89 \\
\hline
 $\omega_{TO,5}$ &663.16 & 667.49  & 669.77 \\
 $\omega_{LO,5}$ &707.60 & 710.03  & 712.00 \\
 $\gamma_{LO,5}$ &21.67  & 13.65   & 10.35  \\
 $\omega_{p5}$   &381.00 & 372.00  & 377.88 \\
\hline
 $\epsilon_{\infty}$ & 4.534 & 4.420 & 4.531\\
\end{tabular}
\end{center}
\end{table}

%
%

\begin{table}
\caption{\label{table2} Fitting parameters for the {\em c} axis reflectance
data utilizing the factorized product form. All parameters are in \cm
except for $\epsilon_\infty$.}
\begin{center}
\begin{tabular}{ccccc}
 Oscillator  & 300K & 200K & 100K& 10K \\
\hline
 $\omega_{TO,1}$ &143.14  &144.04 &144.24 &144.43 \\
 $\omega_{LO,1}$  &147.00 &149.00 &150.33 &149.33 \\
 $\gamma_{LO,1}$  &15.25  &22.55  &18.22  &13.22  \\
 $\omega_{p1}$   &177.06  &203.66 &227.56 &207.74 \\
\hline	
 $\omega_{TO,2}$  &203.00 &203.08 &202.46 &201.31 \\
 $\omega_{LO,2}$  &204.18 &204.65 &204.65 &204.65 \\
 $\gamma_{LO,2}$  &12.00  &12.54  &15.00  &17.00  \\
 $\omega_{p2}$    &147.90 &168.20 &194.94 &243.25 \\
\hline	 
 $\omega_{TO,3}$  &254.16 &255.65 &256.39 &255.25 \\
 $\omega_{LO,3}$  &398.57 &395.57 &397.00 &394.00 \\
 $\gamma_{LO,3}$  &36.22  &29.73  &23.65  &20.88  \\
 $\omega_{p3}$    &988.81 &993.12 &987.34 &982.18 \\
\hline	 
 $\omega_{TO,4}$  &399.33 &396.35 &397.34 &394.61 \\
 $\omega_{LO,4}$  &505.38 &504.45 &502.96 &501.61 \\
 $\gamma_{LO,4}$  &29.96  & 24.39 &19.10  &17.11  \\
 $\omega_{p4}$    &63.32  & 66.10 &43.39  &58.80  \\
\hline	 
 $\omega_{TO,5}$  &526.19 &525.18 &524.13 &523.71 \\
 $\omega_{LO,5}$  &593.24 &595.77 &597.39 &599.08 \\
 $\gamma_{LO,5}$  &41.87  &36.74  &32.82  &32.65  \\
 $\omega_{p5}$   &173.94  &180.02 &185.27 &191.95 \\
\hline
 $\epsilon_{\infty}$ & 4.02 & 4.05& 4.05 &4.05\\
\end{tabular}
\end{center}
\end{table}

\begin{figure}
\epsfxsize=5.5in  \epsffile{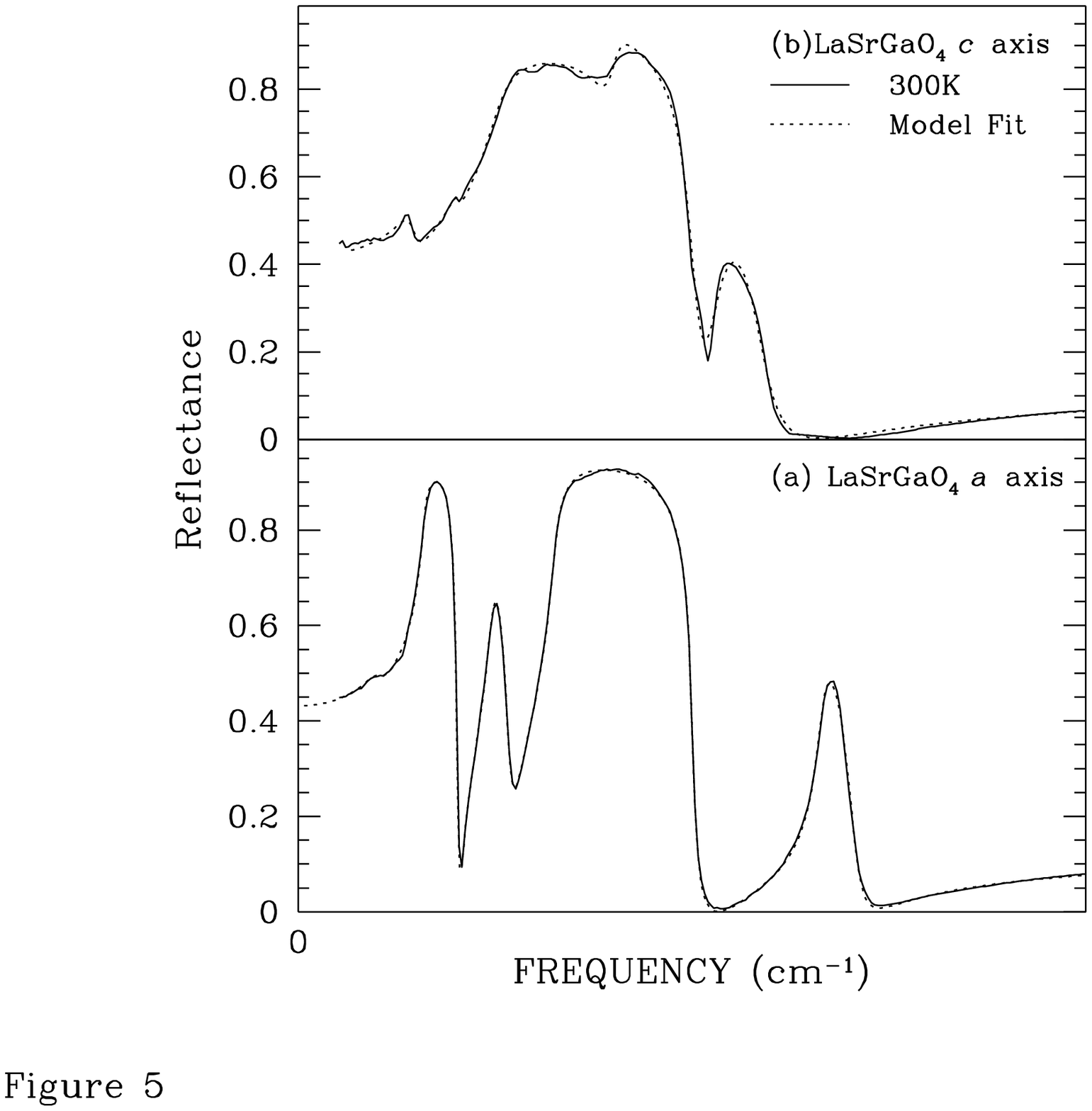}

\caption{\label{rfit} Fits to room temperature reflectance of \LSGO\   {\em
a} and {\em c} axis polarizations. The solid lines are the reflectance data
for 300K and the dotted lines are the fits calculated using values
from Table I and II. }
\end{figure}

The number of phonons used to fit the reflectance exceeds the number
allowed by group theoretical considerations. However, the statistical
nature of the La and Sr placement in the unit cells could result in
additional IR-active phonons due to the broken symmetry.
In particular, the lowest phonon peak in the ab plane fitting is
only included to account for a small shoulder in the reflectance data at
approximately 100 \cm. The remaining 4 phonons result in a good fit to the
reflectance spectrum excluding this shoulder. The {\em c} axis reflectance,
due to its asymmetric peaks, requires 5 oscillators to provide a
satisfactory fit to the reflectance. Figure~\ref{rfit} shows the
reflectance for the room temperature data, both {\em a} and {\em c}
polarizations, and the fits to the reflectance.

%
%

Some mode assignments can be tentatively applied to the fitted oscillators by comparison
to mode assignments for other isomorphic crystals.\cite{burns88,tajima91} For the {\em a}
axis polarization there is one more oscillator than is allowed by group theory.  However
$\omega_{TO,1}$ is a very weak mode and can be assumed to have been introduced by some
symmetry breaking mechanism.  Of the remaining oscillators, $\omega_{TO,2}$ can be
assigned to a mode where the La/Sr atoms move in opposition to the GaO$_6$ octahedra since
the larger mass of the La/Sr would result in a lower frequency of vibration.  The
remaining 3E$_u$ oscillators are vibrations internal to the GaO$_6$
octahedra.\cite{tajima91}

The {\em c} axis polarization has two additional modes above what is 
allowed by group theory. The $\omega_{TO,4}$ mode is
very weak and of the remaining four
oscillators, it is not obvious which one is not an A$_{2u}$ mode. 
It may be that the normally silent B$_{2u}$ mode has been activated. 
The $\omega_{TO,1}$ is likely the mode involving movement of the 
La/Sr atoms with respect to the GaO$_6$ as it has the lowest 
frequency vibration. Similarly the highest two of the remaining 
strong modes, $\omega_{TO,3}$ and $\omega_{TO,5}$ are likely the 
A$_{2u}$ modes since they involve vibrations of the light oxygen 
atoms. This leaves $\omega_{TO,2}$ unassigned. 

In conclusion, we have measured the reflectance of the single crystal,
\LSGO\ , from 50 to 40000~\cm\ for the {\em a} and {\em c} axes including
temperature dependence from 300~K to 10~K. The optical properties are
calculated using Kramers-Kronig transformations on the spectra and the
temperature dependence of the real dielectric and conductivity has been
presented. This allows the dependence of the film spectrum on the substrate
to be determined. This is of importance if the {\em c} axis reflectance is
to be measured on \YBCO\ thin films or other High-\tc\ materials grown on
this substrate.

\begin{ack}

The authors would like to thank C.C. Homes for useful discussions and 
aid in calculations. This work was supported by the Ontario Center 
for Materials Research (OCMR) and the National Science and 
Engineering Research Council (NSERC).  A. McConnell would like to 
gratefully acknowledge the support of B. Clayman and the Department 
of Physics, Simon Fraser University. 

\end{ack}

%
%
%

%
%

\end{document}